\documentclass[11pt]{article}
\usepackage[a4paper,left=3cm,right=2cm,top=2.5cm,bottom=2.5cm]{geometry}
\usepackage{cite}
\usepackage[pdftex]{graphicx}
\usepackage[cmex10]{amsmath}
\usepackage{amssymb}
\usepackage{microtype}
\usepackage{url}
\usepackage{endnotes}
\usepackage{times}
\usepackage{array}
\usepackage[tight,footnotesize]{subfigure}
\usepackage[font=footnotesize]{subfig}
\usepackage{graphicx}
\usepackage{mathtools}
\usepackage{amsmath}
\usepackage{enumitem}

\usepackage{alltt}
\usepackage{enumitem}

\def\boxx{{\vcenter{\vbox{\hrule height.3pt
          \hbox{\vrule width.3pt height6pt
          \kern6pt\vrule width.3pt}\hrule height.3pt}}\;}}

\def\impos{{\;\vcenter{\hbox{\rule[0.15mm]{1.8mm}{1.7mm}}} \;}}
\newcommand{\imposition}[1]{\xrightarrow{~~#1~~~} \kern-10pt \impos}

\def\lrarrow{\leftrightarrow \kern-8pt \rightarrow}

\def\2{\frac{1}{2}}

\def\beq{\begin{eqnarray}}
\def\eeq{\end{eqnarray}}
\def\bex{\begin{alltt}\small}
\def\eex{\normalsize\end{alltt}}

\newtheorem{definition}{Definition}[section]

\newtheorem{proposition}{Proposition}[section]
\newtheorem{unknown-proposition}{unknown-Proposition}[section]
\newtheorem{assessment}{Assessment}[section]

\newtheorem{textpromise}{Promise}[section]

\newcommand{\AOA}{\mathsf{AOA}}
\newcommand{\Def}{\mathsf{Defect}}

\title{Candidate Software Process Flaws for the Boeing 737 Max MCAS Algorithm and Risks for a Proposed Upgrade}

\author{
Jan A.\ Bergstra\\
{\small{Minstroom Research BV Utrecht, The Netherlands}}\thanks{\texttt{janaldertb@gmail.com}}
\\~\\
Mark Burgess\\
{\small{ChiTek-i, Oslo, Norway}\thanks{\texttt{mark.burgess.oslo.mb@gmail.com}}}}

\date{\small{\today}}
\begin{document}
\maketitle

\begin{abstract}
  By reasoning about the claims and speculations promised as part of
  the public discourse, we analyze the hypothesis that flaws in
  software engineering played a critical role in the Boeing 737 MCAS
  incidents.  We use promise-based reasoning to discuss how, from an
  outsider's perspective, one may assemble clues about what went
  wrong. Rather than looking for a Rational Alternative Design (RAD),
  as suggested by Wendel, we look for candidate flaws in the software
  process.  We describe four such potential flaws.
  Recently, Boeing has circulated information on its envisaged MCAS
  algorithm upgrade. We cast this as a promise to resolve the flaws,
  i.e.  to provide a RAD for the B737 Max. We offer an assessment of
  B-Max-New based on the public discourse.
\end{abstract}
\tableofcontents

\section{Introduction}

This paper is a sequel to our introductory paper~\cite{PT4B737}, which
described the unfolding of the Boeing-MCAS affair in the wake of the
tragic crashes of two Boeing 737 Max aircraft\endnote{See e.g.  the
  informative YouTube item about the Ethiopian Airlines flight 302
  crash on March 10, 2019, in
  ~\url{https://www.youtube.com/watch?v=_T5xhHzZjPQ}.}. Below we will 
  abbreviate Boeing 737 to B737. Our tool for
the previous analysis was Promise Theory, and our source information came from
the viewpoint of the resulting public debate.  Unavoidably, as
outsiders to Boeing's software engineering process, we have---at
best---only a rudimentary technical insight into the specifications
for the software component MCAS, and its original requirements. This
limits the kind of reasoning available to us, as outsiders.
Nevertheless, there are still matters which are amenable
to analysis, and may contribute to the public debate.

In this follow-up paper, we have two objectives: to investigate the
extent to which software (or software engineering) flaws can be said
to have played a role in the incidents, and to comment on the
software-related aspects of plans for a redesign.  We make use of
information concerning the B737 Max MCAS algorithm affair, made public
soon after the occurrence of the second disaster, and we consider two
questions: (i) to what extent does the discourse around the incidents
point to possible software design flaws?, and (ii) given that there
are candidates for such flaws, can we narrow the scope of possibility
to be more specific about them?\endnote{In any complex system, like flight
  operations, there is a network of causation that might be ambiguous
  and tricky to decompose.  Nonetheless, certain factors present
  themselves at heart of such a network, and play a decisive role.
  Software is clearly such a issue, as basic and commonly held
  software engineering principles were violated.}.  These
considerations lead us to the notion of a {\em software process flaw},
as discussed in \cite{treatise2}. We refer to such indications as
`candidate software process flaws'.

Without an investigation based on inside knowledge, it's impossible to
accurately determine whether or not these candidate software process
flaws correspond to verifiable problems with the software process that
took place. Thus---as is the case with all public discourse---the
discourse is fundamentally speculative, and conclusions are limited.
Nevertheless, we can sharpen the discussion somewhat.  Johnston \&
Harris~\cite{JohnstonH2019} focus on the question of whether software
engineering can (or must) be considered the cause of the problem.
Their paper provides useful information on the intended working of
MCAS. They conclude that the complexity of the entire problem renders
it too much of a simplification to speak of a software engineering
problem. This is an easy position to take, as being overwhelmed by
complexity is the default state; indeed, the willingness to be satisfied
with such a position might be a result of the popularity of so-called
`blame free' postmortem analysis, and its preference for
pointing away from human factors.  We are unconvinced by that conclusion,
however.  

Promise Theory helps to clarify the roles of intent in a causative
network.  What Johnston \& Harris indicate in detail is that there was
a high risk of feature interaction,\endnote{%
  Feature interaction was noticed as a complicating factor in the area
  of telecom software, see for instance~\cite{KimblerB1989}. Original
  work on feature interaction had a focus on unforeseen and unintended
  phenomena, oversights of designers, whereas more recent work on
  feature interaction has a focus on the handling of conflicting
  actuator inputs (see~\cite{AbdessalemEtAl2018}). In the B737 case
  conflicting actuator inputs were in fact present if, during an MCAS
  intervention, the pilots tried to use the handle on the yoke to
  control the stabilizer pitch.  Following the interpretation of
  feature interaction of~\cite{AbdessalemEtAl2018} the B737 MCAS
  algorithm affair is in part about a potentially unsatisfactory
  solution that was chosen for a specific feature interaction
  problem.}  a suggestion which we have detailed in~\cite{PT4B737}.
Feature interaction, however, is a classic phenomenon in software
engineering and we consider it axiomatic that software engineers must
always be on the lookout for such issues, a task which bestows on them
an overarching responsibility for due diligence.\endnote{%
  We feel that it is not necessarily a service to the discipline of
  software engineering not to be fully open about the possibility that
  software engineering was at fault, just as much as it is not a
  service to piloting not to be open to the possibility that pilot
  errors were involved.  There is a double bind involved, however:
  what may be healthy for the field at large may not be good news for
  some of its specialists.  In other words for an author loyalty to
  software engineering as a discipline and loyalty to its specialists
  may diverge.  We start out with a loyalty to software engineering,
  and a loyalty by default to software engineers, which gives an
  incentive only to accept that software engineering was at fault if
  there are convincing arguments for that hypothesis.  In this paper
  we cannot confirm or disconfirm any hypothesis to that extent, but
  we make some progress towards understanding how that might
  eventually be done.}  Yoshida~\cite{Yoshida2019} claims similarly
that automation is to blame for the disasters, though the paper does
not clearly point out what went wrong in terms of software design or
development. Automation is merely a proxy for human intent, so this
suggests instead that there may have been problems in the intent or
implementation of the automation by software engineers, mixed with many
other issues arising from the intrinsic complexity of Just In Time
design of a B737 NG follow up model. In a Promise Theory analysis,
the role of trust plays a central role: which agents in a process
trusted which others, leading to acceptance of promises being kept
in good faith, and which of those were most vulnerable to being broken?

Choi~\cite{Choi2019} assumes that the software engineering for MCAS,
when considered in hindsight, was indeed flawed or in error\endnote{Again, we use the
definitions of error, flaw, and fault from \cite{treatise2}.}---a viewpoint which
we consider not to be self-evident, and insufficiently supported 
by the remarks made by Choi in his paper. He then takes the case as 
an illustration for the claim that writing safety-critical software 
ought to be considered (legally) as a profession, which
currently is not the case in the US. Choi's paper is obviously 
relevant for our discussion, and we advise any readers who might seek to impute
a conclusion
that any part of our work below is a claim that
Boeing's software engineers may have failed in their due diligence,
to first to read Choi~\cite{Choi2019} in full\endnote{%
  The conclusions drawn by Choi in~\cite{Choi2019} may be
  counter-intuitive for a computer scientist: it is the established
  failure of computer science to provide an empirical or scientific
  foundation for developing safety critical software which apparently
  justifies the transition to a legal regime as it is used for doctors and
  lawyers. Every bug in the software is trivial in hindsight, but the
  fact of the matter is that no software process guidelines have been
  put forward which guarantee the writing of perfect software, just as
  no doctor is able to avoid making mistakes, or to cure all patients.  
  Choi emphasizes that
  writing safety critical software is unlike many other forms of
  engineering where scientific progress increasingly moves what used
  to be professional activity out of the legal regime of judging
  professional malpractice by professional peers, and that for writing
  safety critical software the methodological scientific progress has
  demonstrated that moving in the opposite direction would be
  appropriate.

  The reasoning of Choi~\cite{Choi2019} might, however, also be
  applied to the results in our paper. This works as follows: assuming
  that several convincing candidate software process flaws have been
  spotted, and assuming that further (forensic) investigation would
  confirm that one or more of these software process flaws have
  actually occurred during the construction of the MCAS algorithm and
  the corresponding implementing software component.  Now following
  Choi: a survey of industrial practice would plausibly lead to the
  conclusion that, in most software processes, flaws frequently occur; that
  principled methods for avoiding such flaws have not been successful
  in the past and have not guaranteed the construction of high quality
  software, however defined. Thus, it is only {\em after the fact}
  that these software process flaws have been spotted, and the
  suggestion that successfully avoiding software process flaws is what
  competent programmers can and should do is unrealistic. Software
  engineers deserve to be protected from accusations based on such
  illusions, and that protection is what a legal professional status
  may bring them. The downstream principle   for responsibility, in
  Promise Theory, makes it clear that no upstream promise should be
  taken as gospel under any circumstances.
  
  For the a discussion of the downstream principle for responsibility in 
  Promise Theory we refer to Paragraph 13.6.3 in~\cite{BergstraB2019}.}.

The analysis and discussion concerning software flaws depends on
various propositions, which might be accorded probabilistic
interpretations. This is a common technique in high risk industries
such as aviation and nuclear power. Fault Tree Analysis and other
forms of forensic logic have been in use for decades\endnote{%
  For references to Fault Tree Analysis, see
  \cite{treatise1,faulttree}.  An extensive discussion of propositions
  in forensics can be found in Bergstra~\cite{Bergstra2019}.}.  In
keeping with Promise Theory, and the necessary matching of
propositional (+) and interpretive (-) promises, for every interaction
and observation, we take it as given that a lack of information
confounds the attempt to assign numerical quantifiers, even of
subjective beliefs, within propositions.  All uncertainties must be
taken into account when determining one's subjective belief in a
specific statement. We refrain, however, from asserting 
 any precise probabilities, and use probabilities only
as typical scales about which one may reason by relative orders of
magnitude.  We take these from a semantic set, without specific quantitative
representation: false, very implausible, implausible, fifty/fifty,
plausible, very plausible, and true (verified). This will be sufficient
for our limited purpose.

In the wake of the accidents, Boeing has issued various
announcements\endnote{%
  See \url{http://www.boeing.com/737-max-updates/en-ca/}and the FAQ
  list on that page. From other sources we infer that the system
  architecture will be redesigned in order to have the two on board
  computers working permanently parallel, a modification which as far
  as we understand is unrelated to the MCAS affair, although the
  original alternating scheduling of operation also determines which
  of the two AOA sensors is regularly being inspected by MCAS, a
  mechanism which would then change.} concerning the redesign of the
B737 planes, which it proposes to apply.  We have combined these
snippets of information into a de facto promise named B-Max-New which we
ascribe to Boeing so to say.\endnote{%
  Clearly Boeing did not express any of this in the format of promise
  theory. In fact we express a meta-promise that we consider Boeing to
  have made the promise B-Max-New.} We suggest the idea that the
combination of proposals in B-Max-New constitutes novel subsequent
information which, in a sense of Bayesian statistics can be used as an
incentive to change one's current probability function on the space of
propositions.

\subsection{An algorithm upgrade acquires prominence}

Boeing has announced that---in order to obtain recertification of the
B737 Max---an upgrade of MCAS has been developed which will guarantee
that similar accidents cannot take place in the future
anymore\endnote{This wording is typical in informal usage. In Promise
  Theory, this is only a promise, as there is no such thing as a
  guarantee in practice.}.  Moreover, no hardware modifications will be made to
the B737 Max design beyond providing a standardized warning for pilots
should the two AOA sensors disagree. This warning was an optional
feature previously.  We infer, from these promises, that Boeing has
assessed the problem with the B737 Max to reside solely in the design
of the MCAS algorithm, but not so much in the actual implementation
thereof\endnote{After ignoring pilot errors (including training and
  pilot certification deficiencies), and also maintenance problems
  (each of which which may or may not have causally contributed to
  either of the accidents, what remains is the algorithm's intent.}.
From these statements, one cannot infer that the the MCAS software
component contains a fault or defect (IEEE terminology,
see~\cite{ieeefail}), because that would imply that MCAS had made a
promise that it did not keep. Nor does it follow that the software
process, by which the MCAS software component was developed was flawed
in some way, since it may have implemented the design promises with
perfect fidelity. If the MCAS software component works according to
its promises or requirements, the requirements and promises themselves
may have been at fault. This is a question of whether the software was
a faithful proxy for appropriately represented design intentions.
Following Wendel~\cite{Wendel2019} we use the phrase rational
alternative design (RAD) for a proposed modification of the B737 Max
which would have overcome the difficulties as experienced. A RAD is a
promise, issued by a designer, that upon incorporating the proposed
improvements would result in an alternative design which would not
have suffered from the assumed deficiencies of the design involved in
the incidents.  The problem with third party RADs is that the flight
systems are so complex that one can hardly trust any specific RAD
proposal without actual testing, as the mutual dependencies of system
components are many and intricate.  Thus, as public observers, we may
assess a lack of trust in the ability of exterior promisers to come up
with a RAD for the B737 Max---but we cannot distrust Boeing as much,
by the same argument, since Boeing's engineers have maximal inside
knowledge of the systems. For Boeing, our trust can only be in their
presumed diligence.

\hyphenation{Public}

We interpret Boeing's recent proposed
improvements as a promise (named B-Max-New in section \ref{RAD-B}).
It was made by Boeing, and with the public in scope. This promise
brings new information, which we may use to update our
subjective probabilities, and which we subsequently use as
basis for making two promises about our expectations for the
outcome of the B737 Max improvements:

\begin{textpromise}[A 3th AOA sensor makes trim wheel use less likely] 
From {\bf Authors} (promiser) to {\bf Public} (promisee):

\bigskip\noindent {\sc Promise Body}:
\begin{quote}
Let $n$ be a positive integer $1,2,3,\ldots$
  If Boeing provides an upgrade of the B737 Max, as suggested in
  B-Max-New, then the new model would be operationally closer to the
  B737 NG if MCAS 
  were based on the inputs of 3 (or greater odd number $2n+1$ of) AOA
  sensors, rather than on 2\endnote{This is standard software practice
for quorum determination in redundant algorithms.}.

The use of majority voting for positive sensor outcomes with $2n+1$
AOA sensors would, in particular, reduce the probability that the trim
wheel needs to be used (after a single AOA sensor failure has led to
an AOA sensor disagreement signal to MCAS), instead of the pilot
controlled stabilizer control by wire.
\end{quote}
{\sc End promise}
\end{textpromise}

\begin{textpromise}[Using 3 AOA sensors is ``better''] 
  \label{Max-NewA} 
From {\bf Authors} (promiser) to {\bf Public} (promisee):

\bigskip\noindent {\sc Promise Body}:
\begin{quote}
  If Boeing redesigns the B737 Max, according to what it has promised
  (in our assessment) in promise B-Max-New, then there is a strong
  case that the resulting aircraft warrants pilots to take on
  systematic simulator training, preparing them for the situation that
  MCAS (as upgraded) has been switched off for the rest of the flight.
  Without MCAS active, the behaviour of the airplane does not promise
to be within the constraints imposed by US regulations on avionics:\\
  14CFR \S 25.203(a) ``Stall characteristics''.\endnote{%
    CFR (Code of Federal Regulation), title 14 (Aeronautics and
    Space), part 25 (Airworthiness Standards For Airplanes In The
    Transport Category, including the B737).}
\end{quote}
{\sc End promise}
\end{textpromise}
The rationale for MCAS is rendered less
convincing if extensive simulator training is needed regarding
aircraft control (of the new B737 Max) in circumstances which have no
counterpart in the behaviour of the B737 NG, that is without any
forthcoming MCAS intervention (though with the B737 Max aerodynamics).
Given the need for such training, in any case, one may contemplate 
additional training with a B737 Max configuration but without MCAS
(denoted as B737 Max-min below). Making use of quantitative
AOA information might also provide be attractive to
pilots.\endnote{%
  This option is vacuous, however, if MCAS has an essential role to
  play in obtaining certification as mentioned in
  Promise~\ref{Max-NewA}. Wendel~\cite{Wendel2019} discusses this
  issue in some detail and claims that use of AOA values from AOA
  sensors is common place in military aviation but not in commercial
  jetliners.}
We combine these promises into an implicit one, encompassing the points:
 
\begin{textpromise}[Implausible requirements on simulator training] 
\label{Max-NewB}

From {\bf Authors} (promiser) to {\bf Public} (promisee):

\bigskip\noindent {\sc Promise Body}:
\begin{quote}
If Boeing redesigns the B737 Max, as promised in promise B-Max-New,
then the resulting aircraft is such that either (i) additional
simulator training with MCAS (i.e. its successor) in disabled mode is
essential (for pilots with a B737 NG type rating), or (ii) a further
redesign is required which involves the installation of one or more
additional AOA sensors, in which case simulator training with MCAS
disabled may even become irrelevant, and will become less critical in
any case.
\end{quote}
{\sc End promise}
\end{textpromise}

\subsection{Some propositions}

There are various propositions (hypotheses, assertions, assumptions,
beliefs, and logical variables) which will play a role in evaluating these promises.
It's helpful to introduce some definitions:

\begin{definition}[Positive input to MCAS]
  By this we denote an MCAS event, in which sampled input from one or
  both AOA sensors is considered to be out of range, so that MCAS
  intervention is now required, unless its data set for recent
  interventions suggests otherwise (while the Mach number as estimated
  for the moment is used to determine the strength of the
  intervention).\endnote{%
The Mach number expresses speed w.r.t. air rather than ground, relative to the speed of sound.}  
\end{definition} 

\begin{definition} [The B737 Max-min] 
  This denotes the hypothetical aircraft type which is obtained upon
  disabling MCAS interventions in the B737 Max.\endnote{%
    The B737 Max has no switch which allows pilots to disable MCAS,
    which is understandable from the perspective that pilots are not
    even expected to know of the existence of MCAS.}
\end{definition} 

\noindent Now consider a number of propositions:

\begin{proposition}[C-B737 Max-min certification]
\label{SameTR}
This denotes the assertion that a B737 Max-min class of aircraft would 
obtain FAA certification with the same type rating as the B737 NG.
\end{proposition} 

\begin{proposition} [Cn-B737 Max-min]
  \label{newTR} This denotes the assertion that the B737 Max-min class of aircraft
  would obtain FAA certification, possibly with a new type rating.
\end{proposition}

\begin{proposition}[MCAS not anti-stall] 
  \label{non-AS} The assertion that the B737 Max class of aircraft's
  MCAS software component does not intend to serve as
  anti-stall protection in any way. In other words, switching off
  MCAS 
  would not increase the probability of a stall, compared to the B737
  Max, which would stand in the way of certification.
\end{proposition}
Taking our own position on trust, we shall assume that Boeing engineers have considered and made
up their minds about each of these propositions---perhaps formulated
differently---even though such information is unknown to us.
The information about the B737, collected by Chris Brady on a
website\endnote{%
  Website about the B737 maintained by Chris
  Brady:~\url{www.b737.org.uk}; the information on the site is also
  available in book format.}, provides the following quote about MCAS:
\begin{quote}
  The LEAP engine nacelles are larger and had to be mounted slightly
  higher and further forward from the previous NG CFM56-7 engines to
  give the necessary ground clearance. This new location and larger
  size of nacelle cause the vortex flow off the nacelle body to
  produce lift at high AoA. As the nacelle is ahead of the C of G,
  this lift causes a slight pitch-up effect (ie a reducing stick
  force) which could lead the pilot to inadvertently pull the yoke
  further aft than intended bringing the aircraft closer towards the
  stall. This abnormal nose-up pitching is not allowable under 14CFR
  \S 25.203(a) ``Stall characteristics''. Several aerodynamic
  solutions were introduced such as revising the leading edge stall
  strip and modifying the leading edge vortilons but they were
  insufficient to pass regulation. MCAS was therefore introduced to
  give an automatic nose down stabilizer input during elevated AoA
  when flaps are up.
\end{quote} 
From this, we may conclude that its author would consider
Proposition~\ref{SameTR} (C-B737 Max-min) and Proposition~\ref{newTR}
(Cn-B737 Max-min) to be both false, because the design could not
comply with the regulations, without some form of MCAS.  Avoiding
non-compliance with regulations appears to be sufficient rationale for
introducing MCAS and---whether or not this particular form of
non-compliance would without MCAS, contribute to a risk of
stall---does not really matter for giving a rationale for the presence
of MCAS. We expect, then, that the author of the quote would consider
Proposition~\ref{non-AS} (MCAS not anti-stall) to be true, the
argument being that without MCAS certification will not be obtained,
but not because of the increased risk of a stall, but because of the
invalidity of Proposition~\ref{newTR}.

We maintain that Proposition~\ref{SameTR} implies Proposition~\ref{newTR},
and Proposition~\ref{newTR} implies Proposition~\ref{non-AS}. We
assign the following ad hoc probabilities to these Propositions:
\begin{itemize}
\item Proposition~\ref{SameTR} (C-B737 Max-min) is very implausible
  (if it were true the rationale for MCAS would have been too weak),
\item Proposition~\ref{newTR} (Cn-B737 Max-min) is implausible 
(mainly based on the quote taken from Brady's website), and 
\item Proposition~\ref{non-AS} (MCAS not anti-stall) is very 
plausible (mainly based on repeated information on the matter from Boeing).
\end{itemize}

It should be mentioned, however, that many authors have a different view
on the rationale for MCAS, to mention Johnston \&
Harris~\cite{JohnstonH2019} and Wendel~\cite{Wendel2019} who both
state that the need for MCAS is rooted in anti-stall protection.
We remark that---should Boeing latterly announce that it did indeed understand,
from the outset, that MCAS was meant to serve as an anti-stall system,
then---our trust in Boeing would be decreased, and, as a result, our
odds for Proposition~\ref{PB1} (B-Max-New) would drop significantly.

\subsection{AOA sensor system: false positives versus false negatives}

Let's assume that the AOA sensors, when sampled and the results combined by some algorithm, 
effectively promise one of the following three possible outcomes to MCAS:
``positive'' (AOA too high'), ``negative'' (AOA not too high),
``neutral'' (AOA sensor system out of order). Promise theoretically,
these correspond to a promise for determining the condition for
intervention to be either: kept, not kept, or indeterminate.  We speak
of a false negative if the AOA sensor system (which includes both the
sensor output (+) and its interpretation (-), in promise
terms\cite{BergstraB2014}) produces a negative signal when it should
have been positive. We speak of a false neutral signal if the AOA
sensor system produces neutral value while it should have been
positive.

The idea behind MCAS is principally that, if it receives positive input (AOA too
high), it may impose interventions which overrule pilot
commands, at least initially. Now let us assume that the MCAS
algorithm only intervenes upon a positive input from the AOA sensor
system. In practice neutral and negative are both negative.  We notice
that, if MCAS is meant as an anti-stall system (and thus
Proposition~\ref{non-AS} fails), then (i) MCAS interventions are
sometimes needed because otherwise no certification can be obtained,
and therefore (ii) the occurrence of false negatives and of false
neutrals are mission critical problems.

By minimizing these false negative and neutral assessments, however,
the complementary risk of false positives may increase.  Therefore it
must be promised, at the design level, that MCAS driven interventions,
triggered by false positives will not constitute a significant safety
risk.  This cannot be done by means of an intelligent processing of
AOA sensor information alone; a completely different safety system
would be needed---for instance, by pilots enacting a transition to
manual stabilizer control, using the trim wheel, thereby deactivating
MCAS interventions.

If MCAS does not promise to function as an anti-stall system, false
negatives are unfortunate but manageable, while MCAS interventions
triggered by false positives might constitute a serious risk. That
risk can be dealt with in two ways: reducing the probability of false
positives, by requiring that several AOA sensors are in agreement, or
handing over to the pilot just as in the preceding case, and making
sure, by other design criteria, that MCAS interventions cannot have
adverse safety consequences.

\subsection{How to use two AOA sensors}

If MCAS is meant as an anti-stall system, then it would be natural
for MCAS to intervene if either one of the two sensors notified an
``excessive AOA'', whereas if MCAS is not meant as an anti-stall
system, and false negatives were not so much of a concern, the
prevention of false positives becomes relatively more important. Then one
might then promise that MCAS intervene only if both AOA
sensors agreed about an excessive value. Thus, there is no unambiguous
meaning to ``making
use of both sensors in order to avoid a single point of failure'' if
it has not been determined which failure must be prevented in the
first place.

However, if the method for preventing safety risks from an MCAS
intervention under false positive information is dealt with
sufficiently well (as it must be when it serves as an anti-stall
system), then using the value of a single AOA sensor could only be
defended, if one could disregard the fact that the single point of
failure renders an aircraft system formally `out of order'.
Speculating for a moment that Boeing engineers did not, in fact, know
whether MCAS would be needed for its anti-stall
functionality\endnote{Although the engineers knew that the background
  for MCAS was based on anti-stall, from the Pegasus design.}, and
assigned a subjective probability of `fifty/fifty' to
Proposition~\ref{newTR}, then it could further be argued that the
alternating use of a single AOA sensor makes a kind of naive `Monte Carlo method'
or mixed strategy sense (e.g. as a Game Theory strategy),
because having too many false positives would suggest a need for
additional pilot training on how to use the trim wheel, which
constitutes an option of last resort. This view is hard to defend,
however, as the presence of systems that intentionally limit their own
certainty regarding mission actual information is not the basis for
any kind of engineering standard. Either one assures dynamic stability
without correction, or the envelope for correction should be
dynamically stable.  By not using the disjunction of the ``excessive
AOA'' judgements from both AOA sensors, they made a design decision
consistent with not using MCAS as an anti-stall system, while they did
not fully ``vote against'' that idea either. 

\subsection{White-spots in outsider knowledge}

Regrettably, no perspectives from Boeing software engineers
have percolated to the media, so we lack crucial expert information. 
We do have test pilot perspectives, from scattered emails and quotes that have
found their way to the media as tabloid gossip. Much attention was paid to
the testimony of a Boeing factory `whistleblower', who claims to have
identified problematic working practices and much attention was given
to emails and communications from several Boeing test pilots at the
time of MCAS introduction and tuning. 

For an assessment of the flight system design promises, it seems more
relevant to know what Boeing engineers thought of
Proposition~\ref{SameTR} , Proposition~\ref{newTR}, and
Proposition~\ref{non-AS}, and how they made up their mind on what use
to make of the presence of two AOA sensors for providing input data to
MCAS. As outsiders, we have no clue, and it's conceivable the
engineers came to a fifty/fifty assessment of Proposition~\ref{newTR},
and that the use of a single AOA sensor for each flight emerged as a
convincing compromise.

In Appendix~\ref{APP}, we have made some simplified estimates
concerning the use of one and 2 sensors which illustrate some of the
observations made in the text. We summarize these here.

\begin{proposition}[Deliberate choice for single AOA sensor input]
  \label{cons-single}
  The design decision to take input from a single AOA sensor,
  exclusively, was made. This occurred during the years 2012-2014, in
  a context of deliberation, involving software engineers, airframe
  engineers, and perhaps test pilots. Various alternatives, including
  conjunctive and disjunctive reading of sensor warnings, as well as
  disagreements were balanced as options.
\end{proposition}
We choose to assign the subjective probability ``very plausible'' to
Proposition~\ref{cons-single}, based solely on Boeing's history of
technical standing.

\begin{proposition} [Single AOA sensor input as a compromise]
  \label{cons-single2} The option to take ``AOA too high'' input from
  a single AOA sensor only---and not to require agreement between AOA
  sensors as a precondition---was chosen, because otherwise the
  probability of having to disable MCAS would become too high.
\end{proposition}
We assign subjective probability ``plausible'' to this
Proposition~\ref{cons-single2}. It expresses only the fact that we
have no evidence in either direction. The main virtue of
\ref{cons-single2} is that it summarizes the best argument we can
think of for the decision to use positive input generated by a single
AOA sensor, assuming the truth of Proposition~\ref{non-AS} (MCAS non
anti-stall). We choose to disregard the worst case notion that it
was a result of negligent oversight.

\section{The MCAS affair as a software engineering issue}

Boeing announced an adaptation of the B737 Max
design, which essentially consists of an upgrade for the MCAS algorithm.
From this announcement, we can draw a number of conclusions about Boeing's
beliefs:

\begin{enumerate}
\item Without an upgrade of the MCAS algorithm, no solution to the MCAS
  affair is in sight.  Stated differently, without such an upgrade
  recertification of the currently grounded B737 Max 800 aircraft is
  not conceivable.

\item Both accidents
  may be causally connected to the collective software engineering
  decisions made for MCAS under each set of circumstances.

\item For an outsider to the MCAS software process, there is
  no current evidence that with a different algorithm,
  or with the requirement of supplementary MCAS-aware pilot training,
  that there would have been a different outcome---to the extent that
  either or both accidents would not have occurred.

\item
\label{Softw-Issue}
It's relatively easy to imagine a modification of MCAS, say MCASp
(meaning ``patched MCAS'') which has the property that aircrafts of
the form B737 Max-min[MCASp] would not have become victim of the same
accidents.  In fact, this holds for B737 Max-min. It follows that
merely proposing a modification of MCAS, which would have blocked both
scenarios, comes nowhere near providing a relevant RAD (Rational
Alternative Design).  The question of which role MCAS ought to play
in the first place, has to be taken into account too, before stripping
it of features that may be considered causes of incidents in the
recent past.
	
\item It's unclear to what extent software engineering could indicate
  a level of training needed for pilots, without user experience.  
Software engineering does not promise such assessments.  If training
  methods and efforts, for B737 Max pilots, were considered defective,
  in relation to MCAS, then it would not be natural to consider such a
  deficiency a software engineering procedural failure.

  For the MCAS software component at hand (as in a B737 Max), it is
  entirely conceivable to promise 
  that pilots need not know about its existence in order to fly the
  aircraft safely. From the software engineering perspective, however,
  it is the system behaviour (as perceived by the pilots) which
  matters, not the view of engineers. This is a matter of cognitive
  psychology to determine whether or not some form of knowledge of
  MCAS is helpful or needed for pilots whose entire focus must be on
  quick and reliable determination of which of the available options
  for action need to be taken, and on then acting accordingly and
  decisively (see for instance \cite{jcs}).

\item It is probably the case that fully informed pilot action would
  have increased the chance of survival in the Lion Air case, while in
  the second case a detailed knowledge of MCAS working and timing
  might perhaps have enabled the pilots to recover from what they
  understood as a stabilizer runaway, by making use of the
  control-by-wire of the motor driving the jackscrew, timed so as to
  switch off that motor (and thereby disabling MCAS intervention) just
  before the intervention took place (it seems that they
  tried to do this at some stage), but this ``method'' would be a far
  cry of what the MCAS designers had in mind for the resolution of
  false positive sensor readings, and its successful application would
  have been a highly laudable achievement rather than a reasonably
  predictable professional pilot action\endnote{%
    On 4 October 1992 a Boeing 747-200F freighter aircraft (El Al
    flight 1862) came down in Amsterdam. Years later it was concluded
    on the basis of extensive research that with unconventional
    methods, there are indications that the plane and its crew, not
    tom mention the civilian victims, might have been saved by its
    pilots, Smaili et. al.~\cite{SmailiEtAl2017}. This remarkable
    result does not imply that the pilots should or could have
    discovered a recovery strategy in real time.  The work indicates,
    however, that determination of the flight envelope of a damaged
    aircraft is a highly non-trivial matter, which may produce
    surprising results when pursued systematically.}.

\item Assuming the software engineers were aware that the use of a
  single AOA sensor gave relatively many false positives, implying a high probability that
  pilots would have to fall back on use of the stabilizer wheel, then
  they may yet have had no way of knowing that the stabilizer wheel does
  not so easily serve as a way out of last resort in the absence of
pilot input. It's unclear
  which parts of the inherited B737 NG design the software engineers
  could have expected flawless and unproblematic
  operation, based on their own understanding. This is a ever-present
  risk for engineering about scenarios for which one has no hands-on
  knowledge. By the downstream principle of responsibility in Promise Theory, the aircraft
integrators would have the responsibility to promise this (see \cite{BergstraB2014}).

\item Assuming that the software engineers were aware that MCAS might
  be needed, in the capacity of a safety critical anti-stall system,
  then they had to keep open that option.  In particular they would
  know that false negatives must be prevented (otherwise the system
  might fail to counteract when the risk of stall
  mounted), which may in turn come with an increased potential for
  false positives, and by consequence they would need to rely on the
  assumption that alternative stabilizer control could help with a
  stabilizer runaway. Under these circumstances, it is understandable
  that they did not enforce the nearly full elimination of false
  positives by means of the conjunctive reading of a number of AOA
  sensor outputs (thereby designing as if false negatives don't really
  matter).

\item After the Lion Air accident, it seemed to be the case that
  elementary schooling of pilots would be of great and sufficient
  help. It was only after the second accident that it became clear
  that the presence of MCAS (given the particular features of the
  timing of its interventions) might have somehow undermined the
  capability for the trim wheel to serve as the solution of last
  resort in case of a stabilizer runaway-like problem. It's normal to
  seek to attribute the `blame' of accidents to human error, in first
  instance, since there is always a human at the end of every
  intentional casual chain, and humans are assumed to be less
  predictable components than technological ones. Therefore 
  pointing at pilot error   after the first accident was a plausible idea.

  The situation is even more complicated for the Lion Air pilots,
  however. It   has been claimed that they failed to properly use the manual trim wheel,
  having first switched off both {\sc CUTOUT} switches, in order to deal
  with a problem for which it would have been the appropriate and
  necessary first step.  That observation still leaves the question
  unanswered if---in the Lion Air case---the pilots would have been
  more successful than the pilots were,  had
  they `performed better' according to the designers intended
  script\endnote{%
    A conceptual difficulty in this matter is whether or not repeated
    MCAS intervention based on false positive AOA sensor input may or
    can be classified as a stabilizer runaway for which classic B737
    problem handling procedures apply.}.  A pilot error may have
  occurred, even if avoiding that very error would not have saved the
  plane.  The first accident brought to light the defectiveness of
  B737 Max pilot training, and preparation in certain circumstances,
  but it did not clearly point to a systemic problem in the design of
  the flight control system.
\end{enumerate}

\subsection{Hardware versus software}

We assume that, for many objectives, hardware solutions as well as
software solutions can be developed.  Moreover, we assume that
determination of precisely which issues are delegated to software
takes place at least in part during the algorithm design phase.  The
principle of separation of scales ought to play a role in this, as
well as the assessment of the intrinsic stability (inevitability) of
the processes (algorithms) used in the implementation (see \cite{treatise1,treatise2}).
Consider the following scenarios and design possibilities 
concerning the scope of software control relevant to the discussion:
\begin{itemize}
\item It might have been natural---in a software design process for
  MCAS---to insist on a hardware addition, in the form of a third
  AOA sensor to simplify reasoning. This would be in keeping with `standard
  lore' for quorum systems, where $2n+1$ agents are expected to reach
  a majority outcome, e.g. see \cite{treatise2}.

\item One may imagine that, instead of producing repeated
  interventions, MCAS would switch off itself after two interventions
  and then instruct pilots to engage in full manual flight with focus
  on keeping the AOA within safe limits.

\item One may also imagine that frequent inspection of both AOA sensors
could be integrated with a machine learning system, 
  allowing software to diagnose  sudden anomalous changes in one of the two
  sensors, e.g. by simple machine learning, which---in the absence of
  pilot actions---may be taken for an indication that the sudden
  change was anomalous, and indicating that the other sensor should
  be consulted.

\item One may imagine an MCAS algorithm in which manual use of the
  trim wheel deactivates MCAS, thereby reducing the range of
  conditions which are to be handled by means of the MCAS algorithm.
This approach is used in car cruise control systems, for instance.
\end{itemize}
We don't claim that any of these suggestions would have been helpful
in avoiding an accident, but we find it plausible that the the precise
conditions under which software has dominion are determined, in part,
during software design.  This points to the need for a robust
iterative dialogue between hardware and software throughout the design
of any specialized system---not least one that is mission critical.

\subsection{Criteria for the presence of a software issue}

We now focus on which criteria might qualify the MCAS affair to be
classified as a software or software engineering related incident.
Software must play a role if and only if the following conditions are
met in combination:

	\begin{enumerate}
	
	\item The B737 Max cannot be recertified as is under the B737
          NG type rating (there is a problem, with software or
          otherwise).\endnote{%
          Perhaps the requirements on non-certifiability can or should
          be relaxed by dropping the requirement that it is done
          within the B737 NG type rating. That is a rather strict
          requirement, and it is not self-evident that software
          engineers would have the task to achieve that sort of
          objective.}

	\item The B737 Max-min cannot be certified, not even with a
          new type rating (MCAS serves a necessary purpose, although
          its current form does not solve all problems)\endnote{It may
            be an open question whether or not an update MCASu of MCAS
            can be found at all (i.e. in principle) such that B737
            Max-min[MCASu] (with standard AOA disagreement warning to
            pilots) could be certified by the FAA under the same type
            rating as the B737 NG, where certification is performed in
            compliance with the highest currently available standards
            of certification methodology.  The quality criterion on
            certification may involve improvements on the
            certification process which has been in place for the B737
            Max, but which does not require the introduction of a
            completely novel certification methodology, a path which
            has been suggested in~\cite{Sgobba2019}.  It would be
            unreasonable to ask for a (mathematical) proof that no
            adequate update MCASu of of MCAS can be found, even if
            that were the case.  Indeed there are no practical proof
            systems for the non-existence of algorithms or of
            implementations algorithms.}.

	\item 
          \label{flawSP} Assuming that MCAS development took
          place, within the bounds of a software process model $M^b_{sp}$,
          adopted by Boeing for the design and production of
          avionics control software.  Then either (a), (b),  or (c):

	\begin{enumerate}
	\item There was an error in executing the software process---a 
          convincing indication that the actual
          software process, as it was followed, for the design and
          implementation of the MCAS algorithm, deviated from the
          prescriptions of $M^b_{sp}$ to such an extent that the
          following holds:
	
          Had the prescriptions of $M^b_{sp}$ been followed,
          then either the MCAS algorithm would have become different
          (due to deviating requirements) or else the MCAS
          software component would have been different (as the result
          of more rigorous testing or as a consequence of attempts to
          deliver a formal verification).
	
          In either case a software process flaw has been found (which
          may but need not explain the accidents).  What matters is:
          (i) the resulting MCAS software component was deemed
          adequate, but only on {\em insufficient} grounds, and, (ii)
          that fact could have been noticed had $M^b_{sp}$ been
          properly applied. In this case, a promise being made by
          a low fidelity agent \cite{treatise2}.
          
        \item There was a flawless application of
          software process $M^b_{sp}$, which lead to say MCAS, yet it is
          concluded in hindsight that the MCAS algorithm or the MCAS
          software  component contains an error, 
          which could be repaired once detected, by making a few relatively 
          minor modifications to the algorithm and to its implementation.
          
          In this case it must be analyzed why proper application of $M^b_{sp}$
          failed to detect the error, and why the error was not noticed 
          during certification either.
	
        \item Alternatively, there was a flawless application of
          software process $M^b_{sp}$, which lead to say MCAS
          and in hindsight it is concluded that MCAS does not solve the
          engineering problems so that in any case MCAS must be redesigned,
          though without the guarantee that such a redesign can or will be successful.
          	
          In this case, a rather fundamental flaw in the software process model has been
          discovered and a modified software process $M^b_{sp}$ must
          be promised first. Moreover it is not clear whether or not
          an upgrade of the software component MCAS can be found at
          all (because, unfortunately, it has come about that a
          ``correct'' process for the development of such a solution
          within $M^b_{sp}$ does not conclusively prove the existence
          of an outcome that can keep all promises.)
	\end{enumerate}
	\end{enumerate}
	We have no reason to assume (b), and the presence of a straightforward 
	implementation error (in a context with adequate specifications for MCAS) 
	would certainly have reached the media by now. Moreover, 
	we have no means to establish (c)  even if that were the case,
        and we limit attention to options for software process flaws
        as classified under (a).	
\subsection{Candidate software process flaws}  

From our vantage point, it is impossible to determine whether or not
the software process for MCAS contained flaws, as referred to under
~\ref{flawSP} (i) above.  What we can do, however, is to suggest some
candidate flaws.  In this Section we propose four candidate software
process flaws\endnote{%
  Some software process flaws, candidate or actual, clearly don't
  matter for the case at hand, for instance if the testing coverage of
  a software component taking care of unfolding the landing gear is
  considered insufficient, there is a process flaw, even if no defect
  has been missed which might have been detected by means of a check
  against a proper test suite, but such a flaw cannot possibly have
  any relation to the two accidents from which the MCAS affair has
  arisen.}.

\subsubsection{Questionable safety level classification}

This section proposes, by a hypothetical argument, that the acceptance
of a safety certification assigned to MCAS was itself a design flaw
that may have warranted further attention.

As reported, in various documents, the MCAS software component
has been assigned a {\bf DO 178c} safety level B (deemed `hazardous').
It may be the case that no convincing scenarios have been provided
which support this classification.  In that case, one candidate
process flaw consists of insufficiently matching high level
requirements (e.g. anti-stall) and low level requirements (e.g. increase
counter when event received) on the MCAS algorithm.
A justification for this could go as follows. If MCAS were meant to serve as an anti-stall
system, and if an MCAS failure would somehow prevent it from playing
that role, a stall might occur which may be catastrophic (hull loss)
rather than than hazardous (risking the lives of some passengers or
crew members at worst).

It seems to follow that the level B assignment corresponds to not
taking MCAS for an anti-stall system, so the question arises which
other scenario could justify classification B. Now assuming that the
main virtue for MCAS is supposed to be to make the B737 MCAS
maneuvering characteristics similar to those of a B737 NG, it is hard
to imagine that failing to play that role could cause a hazardous
outcome (for instance destabilized by sudden turbulence, or side wind,
which risks lives of passengers and crew not wearing seat belts).

The safety classification `hazardous' (assuming it had been justified
in a systematic manner) thus relates apparently to the handling of
MCAS interventions, which may have been based on false positive
information.  Adopting that assumption, however, means nothing less
than that it was known, at design time, that pilots handling one or
more MCAS interventions were confronted with a level B safety risk.
Now, the rationale of the MCAS system becomes doubtful:
the cure is worse (in terms of its safety risk) than the problem it is
supposed to solve (again as measured in terms of a safety risk).

The above argument does not change much if MCAS were given safety
level A (potentially catastrophic).  It seems that the only reasonable
safety certification one could accept of a properly functioning MCAS
component would have been C (major problems) or below.

\subsubsection{Implicit awareness of the state of the art} 

The question of whether or not MCAS is supposed to serve as an an
anti-stall system may have been on the table, even though undetermined.
In any case, the experience and
knowledge associated with its Pegasus predecessor may have been
insufficiently acknowledged.  In this case, it's arguable that not
enough use was made of the existing experience with the MCAS high level
requirements specification, inherited from the Pegasus project. This
specification probably has been imported in the B737 Max software
process.
 
\subsubsection{Unfamiliar requirements} 

 \label{equation}
 Assuming that MCAS is supposed to make sure that say a B737 NG is
 ``pilot experience equivalent'' to
 B737 Max, that problem amounts to solving the informal promise equation:\\
 
 $~~~~~$B737 NG $\equiv_{pe}$ B737 Max-min[X] \\
 
 \noindent where $\equiv_{pe}$ signifies equivalence, or the
 indistinguishability of promises assessed by an agent for the left
 and right hand sides.  Here X is a possibly multi-variate variational parameter, for a software
 component, for which MCAS is claimed to constitute a solution, and X
 implements a thread (i.e. is the mathematical semantics
 of an instruction sequence), which is combined in parallel with the
 threads of B737 Max-min by way of strategic interleaving.\endnote{%
   We refer to~\cite{BergstraM2007} and~\cite{BergstraM2008} for
   multithreading with strategic interleaving.}  Now, in this case,
 the candidate flaw is that assessing high level requirements depicted
 in this this equation is not currently possible---it can be
 considered too remote from what computer science has on offer, in
 terms of languages and concepts capturing requirements. Formally
 defining $\equiv_{pe}$ is very difficult. Aside from Promise Theory,
 which can address a scaling of issues and assess their gaps and
 flaws, and yet is still insufficient, we are not aware of papers on
 this or similar issues.  Then there is the question of assessment,
 which may be performed differently by each assessor. Thus, it's
 unclear how compliance with such high level issues could be
 demonstrated at all. This is a major unsolved issue in computer science.
 
 To put it another way, if a set of promises (perhaps arising from a
 requirements specification) were the starting point of MCAS
 development, then there is no evidence that merely applying a
 traditional software engineering process (such as {\bf DO 178c})
 could produce such an $X$. Software engineers should not have
 accepted the assignment in this case, as it would be clear that the
 promises could not be kept.
 
 \subsubsection{Intervention overshoot} 

 The MCAS algorithm commands highly disruptive interventions, in terms
 of the size and the duration of the movement of the stabilizer. The
 candidate flaw is that not enough research and development were performed to justify
 the quantitative aspects of MCAS interventions.
 This candidate flaw is connected with the assessment of the safety
 level of MCAS. The stronger and longer its interventions the higher
 the risks connected with interventions triggered by false positive
 sensor inputs.

\subsection{Algorithms: a definition}

\label{AlgoDef}
The concept of an algorithm is becoming more important day by day, but
there is no consensus on what it means.  Algorithms are now the
subjects for public debates, e.g in facial recognition, game playing,
and of course in flight systems. The rules of engagement for such
public debates cannot simply be inferred from a scientific tradition
in computer science, or from a practical tradition in software
engineering.

A difficulty faced by a public debate about any specific algorithm
comes from the fact that none of the participants are likely to be
aware of the interior details of the design of the algorithm, nor with the
specifics of implementations of it. Both in computer science and in
software engineering, this removal between the actors in a discourse
from the technicalities of its subject matter makes discussions rest
on speculative and principled arguments rather than facts.
It seems to have gone unnoticed by the computer science community that
the word `algorithm' serves to decouple software technicalities from
their underlying ideas, so as to make implementations amenable to
public debate.  This aspect constitutes part of our definition of an
algorithm for the purposes of this paper:

\begin{definition} [Algorithm]
\label{defAlg}
An algorithm is a method for solving a certain problem (i.e.
for achieving a promised outcome) in a stepwise manner\endnote{Not all problems may 
necessarily be amenable
to this mode of solution, but we shall assume they are in this case.}.
An algorithmic method is documentable by a finite sequence of
instructions\endnote{For the notion of an instruction sequence and the
  consequences of requiring finiteness thereof we refer
  to~\cite{BergstraM2012}.  Our definition of an algorithm is somewhat more
specific than most definitions conventionally used in computer
science, e.g. in Corman et. al.~\cite{CormanLRS2009}.}.  An execution of the method implies
performing the instructions, possibly more than once, so that no
{\em a priori} bound on the number of steps can necessarily be given in terms
of the number of instructions.  Algorithm refers collectively to this
method as a conceptual black box or semantic boundary, which makes
certain exterior promises, without having all interior details at
hand\endnote{This is what is known as the separation of interior and
  exterior promises in Promise Theory.}.  The latter understanding of
algorithm renders it amenable for debate among software
non-specialists, on the assumption that it keeps its promises.
\end{definition}
We assume that the concept of a public debate about an algorithm and
its implementation is reasonable, and that its implementation as a software component makes
sense, as these represent intentions, which may have societal and even
legal ramifications.
Our discussion, as part of such a debate, is motivated by the question of whether
the design and implementation of the MCAS algorithm may have contributed causally
to the fatal accidents, and we proceed exclusively with the handicap
of information pertaining only to the public debate on the matter.
Our ruminations are therefore likely to be of particular interest to
lawyers and public officials.

\section{A RAD for MCAS and a potential candidate software process flaw}

\label{RADflaw}
According to Wendel~\cite{Wendel2019}, if a litigation lawyer plans to
make a case concerning a suspicion of a design problem in a product, it's a
wise idea to provide a so-called RAD, or a Rational Alternative Design.
The RAD indicates how the design problem could have been avoided with
the implication that it should have been avoided. Indeed many authors
have made suggestions on how the MCAS system might alternatively have
been designed. We take the following position, however: the complexity
of flight control software is so high that it is not plausible for an
outsider to suggest an RAD because too many factors must be taken into
account.
However, a RAD for the B737 Max which has been proposed by Boeing
cannot reasonably be dismissed as potentially not taking the
complexity of the problem into account---since they are the
insiders we seek.  Therefore, it's reasonable to look into such an RAD
(if available) in some detail\endnote{%
  It would be unreasonable to hold the existence of an RAD emerging
  from Boeing against Boeing, because it can hardly be assessed by an
  outsider how difficult it has been to come up with and to validate
  its practical feasibility.}.

\subsection{An RAD originating from Boeing}

\label{RAD-B}
By RAD-B, for the B737 Max, we denote the following outline:
\begin{itemize}

\item Improvement of various software components (different from MCAS), as well as further development of the computing architecture within the aircraft in such a manner  that in the future both main flight control computers will permanently work in parallel,
rather than that in an alternating fashion one of these is in the lead during a flight, as was the case until now in 
the B737 NG 800 and the B737 Max 800.

\item The MCAS algorithm is modified, and accordingly the MCAS
  software component is upgraded, resulting in MCASu in our
  terminology.

\item The upgrade of MCAS to MCASu is expected to enjoy the following
  properties:

\begin{itemize}
\item Upon discovery of an AOA sensor disagreement, the MCASu
  component will be terminated and act as if no ``AOA too high''
  triggers are being received until reactivated.
  In addition the pilot will be expected to do without autopilot, without
  autothrottle, and without `control by wire' of stabilizer trimming,

\item Interventions by MCASu will be more moderate (changing the trim
  less) than before,

\item MCASu interventions won't be repeated (within the same intervention episode), i.e. unless in between an episode with adequate AOA and normal flight parameters has occurred (i.e. has been observed),

\item MCASu will allow overriding (or rather, successful
  counteraction) by ordinary pilot control via the yoke.
\end{itemize}

\item Each cockpit will routinely be equipped with the same visual
  and acoustic signal for warning in case of an AOA disagreement,
  (this upgrade constitutes the only hardware modification, and was
  already available as an optional feature\endnote{%
    Wendel~\cite{Wendel2019} provides convincing arguments why the
    feature has been offered by Boeing as an optional one.}).
\end{itemize}
Now Boeing has issued a promise, B-Max-new, as follows:

 \begin{textpromise}[B-Max-new]
 \label{PB1}
\bigskip\noindent
 From promiser: {\bf Boeing} to promisee: {\bf Public},

\bigskip\noindent {\sc Promise Body}:
\begin{itemize}
\item A redesign of the B737 Max has been realized (we, not Boeing,
  call it B737 Max-min[MCASu], the B737 Max with MCAS upgraded to
  MCASu),
\item Many test flights have been made with the available B737
  Max-min[MCASu] prototypes, and simulator sessions were carried out
  with a novel dedicated simulator for the B737 Max-min[MCASu],

\item The B737 Max-min[MCASu] will obtain within a reasonable time the
  FAA (re)certification for the same type rating as the B737 Max had
  been certified with (and the same as the B737 NG).

\item Preparing existing aircrafts to the mentioned upgrade is likely to  
involve not more than 150 expert working hours per aircraft.
\end{itemize}
{\sc End promise}
\end{textpromise}
From B-Max-New the following conclusions can be drawn:
\begin{itemize}
\item Boeing engineers assume the truth of Proposition~\ref{non-AS}
  (MCAS not anti-stall). A reasonable consequence of taking notice of
  the announcement of B-Max-new is to raise the subjective probability
  of Proposition~\ref{non-AS} (MCAS not anti-stall) to ``true''.

  If one of both AOA sensors is damaged during the start, say by way of
  collision with a bird, the MCASu system will be disabled so that MCAS
  cannot provide any protection against stall.  As the risk of a
  single sensor failure is relatively high, the importance of MCASu
  for anti-stall protection must have been considered minimal or
  absent by Boeing engineers (and so may we).

\item The MCASu feature is considered useful, although it will be
   less intrusive than its predecessor MCAS. This makes one
  wonder why the design decisions for the MCAS algorithm were
  as they were.  The question arises whether the
  functionality of MCASu could be specified in simple terms, for a large
  audience, now that the suggestion that its claim of anti-stall functionality is
  conclusively outdated.
\end{itemize}

\subsection{Assessing RAD-B and B-Max-New}

 Our assessment of Promise~\ref{PB1} (B-Max-new) is:

\begin{assessment}[B-Max-new]
  Boeing may not be able or willing to keep promise (B-Max-new) and and
  therefore we qualify our assessment of Promise~\ref{PB1} (B-Max-new)
  as implausible.
\end{assessment}
We shall argue in a number of steps, while assuming significant trust
in Boeing to begin with.

\paragraph{Step 1.} We remark first that without trust in Boeing there
no reason to believe that the promise of (re)certification of the B737
Max-min[MCASu] will be kept at all. The challenge is then to justify a low
probability assessment of promise~\ref{PB1}, while maintaining significant trust in
Boeing.

\paragraph{Step 2.} Given the assumption of significant trust in Boeing,
our assessment of promise 3.3 (non-anti-stall promise) in~\cite{PT4B737}
is that it will be kept. If follows that MCAS is not helping out in
life threatening situations but only in cases where a mismatch (in
terms of expected pitch stability according to Boeing) with pilot
experience with a Boeing 737 NG can result.

\paragraph{Step 3.} 
If Proposition~\ref{SameTR} (C-B737 Max-min) were the case, the
introduction of MCAS can hardly be justified, the only remaining
relevance of it being that it makes the B737 Max pilot experience so
close to the B737 NG pilot experience that not only the same type
rating as for the B737 NG can be used but also, and more importantly,
no costly training is needed for B737 NG pilots who plan to pilot a
B737 Max. So we find that C-B737 Max-min is probably false.

\paragraph{Step 4.} 
Further from promise 4.1 of~\cite{PT4B737}, as well as our trust in
the promiser (Boeing) we infer that Boeing expects the B737
Max-min[MCASu] to have a single minor hardware improvement only
(except for the modernization of its computing platform with both
computers permanently active and working in parallel, an overdue
innovation which is unrelated to MCAS): the pilots will receive a
uniform standard warning of AOA disagreement of such a disagreement is
noticed, upon which MCASu will disengage and pilots are supposed to
perform manual (non autopilot flight).

\paragraph{Step 5.} With two AOA sensors on board the likelihood of an
AOA sensor disagreement is maximal.  Hundreds of such events have
been observed in recent years. It takes only a bird to collide with
one of both sensors and the subsequent presence of an AOA sensor
disagreement is plausible.

\paragraph{Step 6.} Upon having received an AOA disagreement warning,
a B737 Max-min[MCASu] behaves just as a B737 Max-min until the end of
the flight.

\paragraph{Step 7.} Now Boeing will argue that for piloting a B737
Max-min[MCASu], just as they did before for the B737 Max, that no
substantial further training (either in theory or in the simulator) is
required. From this it follows that, without further training B737 NG
pilots can fly a B737 Max-min, for almost a full cycle (a collision
with a bird may well take place just after lift off.)

\paragraph{Step 8.}  We conclude that the circumstance in which pilots (or prospective pilots)
of a B737 Max-min[MCASu] have received an AOA warning constitutes an
emergency---a high stress cognitive burden. Prospective B737
Max-min[MCASu] pilots will require a meaningful training for that
particular course of events, and moreover they will require additional proactive
simulator training on how to fly the B737 Max-min (i.e. how to fly the
B737 Max-min[MCASu] by hand).
They will likely find it hard to accept that MCAS (MCASu) is needed for their
well-being, as pilots, while its absence creates no difficulty worth
thinking through in detail or worth being tried out in a simulator.
These paradoxical positions cannot easily be explained to the public:
after MCAS has been upgraded (to MCASu), pilots must be specifically
trained in how to work without MCASu.

A further modification of the B737 Max-min [MCASu] which simplifies
these matters is in order.

\paragraph{Step 9}
Having three AOA sensors, instead of 2, with majority voting (the A320
architecture for that matter) constitutes a design which we will denote as B737
Max-min [MCASu/3mv], or even five AOA sensors (B737 Max-min
[MCASu/5mv]) will drastically alter the picture.

In a B737 Max-min[MCASu/5mv] plane, the probability of an AOA mismatch
which disables MCASu, is very low (three AOA sensors must malfunction
simultaneously), and for that reason piloting that plane comes much
closer to handling a B737 NG than is the case with a B737
Max-min[MCASu/2mv] i.e. a B737 Max-min[MCASu] (at least under the
assumption that MCASu works well in the absence of false positives
while not being hit with false negatives too much).

\section{Why does it all matter, and when?}

It's a challenge to express in a reliable manner when,
how, why, and for whom the observations made in preceding sections may
be of relevance. As authors, we must define our own role. We view our
role as that of limited `experts', in a particular field of automated systems and
reasoning, who provide knowledge about a case which may be of use to
all persons and organizations who take an interest in the affairs
concerning the B737 Max. Our combined expertise comprises: (i)
practice and theory of software engineering, (ii) Promise Theory, as
used as a tool for the description of complex multi-agent scenes,
(iii) Logic, including theoretical aspects of forensic reasoning. Our
expertise, however, does not include the following areas: (i) aircraft
engineering, (ii) the actual writing of safety critical software,
(iii) working within the {\bf DO 178c} software process framework,
(iv) piloting, (v) airline management.  Readers may derive an initial
level or trust in us---the authors---from this brief declaration of
claimed expertise. Indeed, readers are free to assign ad hoc
probabilities, as we have, if they so desire.

\subsection{Potential relevance of both forecasting promises}

The potential relevance of the forecasting promises,
Promise~\ref{Max-NewA} (using 3 AOA sensors is better) and
Promise~\ref{Max-NewB} (implausible requirements on simulator
training) is now as follows: these may be taken for our promises on how
the acceptance may work out of an aircraft model B737 Max-new,
engineered along the lines of Promise~\ref{PB1} (B-Max-new). In
particular, pilots (and airlines) may insist on rigorous simulator
training on the basis of a positive assessment of
Promise~\ref{Max-NewB} and they may insist that simulators represent
the various scenarios for using the manual trim wheel very well, and
may ask for evidence for the latter.

A positive assessment of Promise~\ref{Max-NewA} (using 3 AOA sensors
is better) and the arguments given for it, may produce an incentive
for airlines as well as for some members of the public to insist that
more than 2 AOA sensors will be installed in a renewed B737 Max model,
and to insist that Boeing won't keep Promise~\ref{PB1} (B-new-max),
and instead will withdraw that promise and issue an upgraded Promise
including one or more additional AOA sensors.

\subsection{Potential relevance of the candidate software process flaws}

Adequately analyzing the potential merits of pointing out candidate
software process flaws, for the software engineering process which led
up to the MCAS algorithm, is significantly more difficult than the
argument just given, in cases of the Promises~\ref{Max-NewA} (using 3
AOA sensors is better) and Promise~\ref{Max-NewB} (implausible
requirements on simulator training).
We shall distinguish five different lines of discourse, for the appreciation of the potential
relevance of this work, each of which may be relevant for some
readers, and may be irrelevant for other readers.

\subsubsection{Curiosity}  A general and impartial curiosity drives interest
in possible explanations for the tragic incidents. Our work may contribute to that
perspective by providing hypothetical candidate explanations in some
detail.

\subsubsection{Litigation support: top level division of responsibilities}

\label{agreement}
Victims and stakeholders in the crashes may intend
to sue some combination of an airline, the airframer, suppliers of the
airframer, one or more certification authorities. Supporting these parties is another line
of interest. A fuller picture may enable
the plaintiff or sued parties to an agreement, to share
responsibility for the problem each of them will accept, or reach a
corresponding division of claims.
It may be useful for all sides involved to have a common language for
talking about what might have gone wrong with the software engineering,
even if parties agree not to investigate in detail what actually went
wrong.
\subsubsection{Litigation support: if details on software engineering matter}

Several different scenarios may involve a setting where the details of
software engineering do matter. If Boeing, airlines, certification
authorities, and perhaps other parties cannot arrive at an agreement
about a division of responsibilities, as mentioned in
Paragraph~\ref{agreement}, then they may want to settle the matter on
the basis of a detailed investigation of facts so that many more
details of organization and production come into play.

In a different scenario, a part of the software process has been
outsourced to one or more suppliers and a division of responsibilities
must be found for the parties involved in the relevant software
process. Then it may matter for all sides to have a survey of candidate
software process flaws at hand.

\begin{itemize}
\item By using the language of software process flaws, rooted in
established Promise Theory, we provide a way
  to get around the straightjacket of having to suggest a RAD, which
  may then be easily judged technically naive by the
  defendant's lawyers.

\item If the framework and the logic of software process flaws has
  been accepted by various sides in principle, then the candidate
  flaws point at options for the prosecution or for the claiming side
  in a civil case.

  By accepting said framework, including a survey of candidate software
  process flaws, a court (providing it uses Bayesian reasoning with a
  subjective understanding of any assigned probability) also promises to adopt (and
  keep fixed for the time being) non-zero (undisclosed) prior odds for
  each of the listed candidate software flaws. The choice of such
  prior odds is entirely left to the discretion of the court. By
  adopting non-zero prior odds the court opens a potential path
  forward to proof of existence in case no decisive and confirmatory
  testimony from software engineers who participated in the process
  can be obtained.

\item As it stands the four candidate software process flaws
  constitute mere speculation and carry no weight of proof. Evidence
  may be gathered, however, first of all by means of interviewing
  relevant software engineers (if they are willing to cooperate).

\item In lieu of confirmation, for the mentioned software process flaws, 
the claimant may need
  to use indirect proof. In an indirect proof, say for an occurrence of the
  second software process flaw 
  (wrong assignment of safety level to a software component) a certain type of
  evidence $E$ is brought forward (by the claimant or by the
  prosecution, the information resulting from forensic investigation)
  and the validity of $E$ is confirmed by witness testimony. Moreover,
  in advance of the testimony and independently of that testimony a
  software process expert (not one of the authors, however, as we both
  don't possess the required expertise in software metrics) have been
  able to determine a significant likelihood ratio from which it
  follows that the evidence $E$ was much more likely to hold true in
  the presence of the second software process flaw than in its absence.

  As stated, the court may adopt non-zero prior odds for each of the
  listed candidate flaws, including the third one.\endnote{%
    A growing base of documented experience is emerging from DNA
    related cases where likelihood transfer mediated reasoning is used
    quite often (see also~\cite{Bergstra2019}).}  Now the mentioned
  software process experts (in general, not specific for Boeing
  software engineering) may communicate the likelihood ratio which
  they have determined to the court from which it may derive (compute)
  its posterior odds, on which a final decision may be
  grounded.\endnote{%
    The actual values of prior odds and of posterior odds as used and
    computed in a case, as well as height of the threshold for the
    posterior odds which is used by the court to arrive at a discrete
    decision (yes or no), is all private to the court and will not be
    communicated to any other participant in the process. The
    likelihood ratio, however, as communicated by the experts to the
    court is included in the formal proceedings of the case.}
\end{itemize}

\subsubsection{Software process improvement}

Engineers may seek to improve  the
execution of the software process, or in case no single software
process flaw can be identified retrospectively, they may pursue the more far
reaching objective to improve the software process at hand.

\subsubsection{Towards responsible software engineering}

A requirement, which might be imposed in the long run for
MCASu, is that it will be the outcome of so-called responsible software
engineering (RSE). A criterion for RSE as mentioned in
Schieferdecker~\cite{Schieferdecker2020} is explicability, which goes
beyond the narrow circle of software technicians involved in a
project.\endnote{%
  In~\cite{Schieferdecker2020} software is to be defined by so-called
  meta-algorithms, a role which we ascribe to algorithms proper. We
  mention that (in~\cite{Schieferdecker2020}) following ISO
  terminology software is considered an intellectual creation which is
  independent of the medium on which it is represented.  Following our
  definition of an algorithm (see Paragraph~\ref{AlgoDef}), the
  algorithm is an intellectual creation which is even more abstract
  than software components (a part of software), and more abstract
  than the documentation of software components (another constituent
  of software) it is independent from the precise form of its various
  documentations, and for that reason and by definition not a part of
  the software.} We expect that thinking in terms of the removal of
software process flaws may be helpful for obtaining explainable
algorithms.

\section{Concluding remarks}

In this paper we have obtained three results: (i) criteria under which
the B737 Max MCAS algorithm affair may be considered a software issue,
(ii) a list of four
so-called candidate software process flaws each of
which might explain in part what went went wrong in the B737 Max MCAS
algorithm affair (questionable safety level classification, implicit awareness of the state of the art, 
unfamiliar requirements, and intervention overshoot), and (iii) an argument why the recent promise by
Boeing about the alternative design of the B737 Max may require a
further revision.

\subsection{MCAS design as a case of DSR (Design Science Research)}

Concerning the design of MCAS---and now of MCASu---one may notice that,
from the outset, the very existence of a software component X which
solves the equation mentioned in Paragraph~\ref{equation} is an open
question. It follows that there may be none or perhaps many solutions
to this problem. It also follows that we are faced with a research problem to
better understand what we might call Design Science Research (DSR), i.e. ``how to solve
a problem by implementation of a process leading to an innovative artefact''
for which each of the 7 guidelines listed in Hevner et.
al.~\cite{HevnerMPR2004} are relevant. The (re)engineering of MCASu
might conceivably be cast in the framework of DSR. Doing so would
focus attention on guideline 6 (design as a search process) which
calls for a thorough understanding of the relevant search space and
for the application of a systematic method for selecting a solution
within the given search space.

Constraints on the search space just mentioned may be split into
intrinsic constraints and extrinsic constraints. Intrinsic constraints
guarantee safety and compliance with strict rules of airframe
behaviour (if any such rules outside the realm of safety are being
imposed), while extrinsic constraints are introduced in order to
optimize physical pilot experience, amenability to simulator based
training, compatibility with the minimization of memory items,
passenger experience, and explicability to prospective customers.
Expanding on external requirements, one may imagine that an MCAS
successor system includes assistance for the pilots when its
interventions take place and pilots may feel the need to intervene
nevertheless with their own interventions.  The terminology of
intrinsic requirements and extrinsic requirements has been elaborated
in detail in the context of care robots, see e.g. Poulsen, et.
al.~\cite{PoulsenBT2018}.

Finally, MCASu may be considered a software robot which can therefore be thought of in
terms of robotics and AI.  The ethical issues arising from this are
a separate issue, and Poulsen
at. al.~\cite{PoulsenBT2018} suggest that a ``school'' must be chosen
concerning machine ethics. There are some options: (i) moral
(software) robotics in which decision-taking is done by the robot
which is given ethical principles from which it must compute adequate
decisions, (ii) good robotics, an approach in which the designer views
the (software) robot as an ordinary program which must be ethically
adequate by design, (iii) a combination of the two where the system
implements moral robotics, while performing as if it were designed as
an artefact for good robotics.  The combined approach appears in a
different terminology in the context of autonomous weapons, e.g.
in~\cite{ElandsEtAl2019}

\subsection{From secrecy about the presence of a software component to secrecy about its\\internals}

It has been stated, in many blogs and papers on this affair, that Boeing was negligent
in not disclosing the existence of MCAS algorithm and its
implementation in a software component MCAS to pilots. For instance
Hutton \& Rutkowski~\cite{HuttonR2019} qualify the secrecy about the
presence of MCAS as implausible.  Obviously, when announcing the
existence of the MCAS software component to customers, pilots, and
perhaps the public, some abstraction of its algorithm must be
communicated, as the whole algorithm may comprise too many
instructions for a non-(software)technical audience to swallow. In
this case secrecy is a matter of degree and is plausible that some
parts of the MCASu algorithm will remain unexplained to pilots and
customers.\endnote{%
  In Bergstra~\cite{Bergstra2019b} the number of instructions of an
  instruction sequence $X$, denoted LLOC($X$) for ``logical lines of
  code of $X$'' is used as a metric for its size.}

If there is any principle which applies to the amount of information
conveyed to pilots, it appears only to be that such information
be minimized rather than maximized.  It seems still to be a useful
design requirement on MCASu that its very existence need not be
explained or announced to pilots, if only as an ideal which for other
reasons won't be achieved in this particular case.  Forgetting for a
moment that it has become impossible to hide existence of MCASu after
the existence of MCAS has been made public, it seems to be the case
that accepting the need to explain the working and the existence of
MCAS (MCASu) to pilots goes hand in hand with acknowledging its
potential deficiencies.\endnote{%
  A definite need to be open about the existence of MCASu would arise
  if the idea is that pilots would apply some form of reinforcement
  learning in order to minimize the number of MCASu interventions on
  the long run.}

MCASu might become more complex in the future as its presence turns
the plane in part into an ``Autonomous Thing (AuT)'', making its own
decisions to intervene during a flight. It may in due time be equipped
with learning capabilities, and be made adaptive to the operational
style of a certain pilot. Following Linz~\cite{Linz2020} an MCASu with
learning capability must not be safety critical, however, at least not
for the foreseeable future, and from this it follows that the degrees
of freedom for its design are limited by the safety level
classification which is assigned to it.

\subsection{Addendum: Promise Theory}

For expositions on Promise Theory and applications thereof we refer to
Burgess~\cite{Burgess2015} and Bergstra \&
Burgess~\cite{BergstraB2014, BergstraB2019}. In Bergstra \&
Burgess~\cite{PT4B737} an extensive introduction to the connection
between Promise Theory and the rather singular technical MCAS affair
is provided.

\subsection{Disclaimer}

None of our remarks are based on new evidence. We work solely with
publicly available information only as a principle of methodology.
The validity of the information we have used cannot be taken for
granted.  We combine available information into patterns and, in that
manner, try to extract meaning out of disparate snippets of promised
information. We believe that reasoning about technology on the basis
of incomplete, and sometimes invalid, non-technical and publicly
available information, has become important, and requires the
development of its own methodologies. Our ``promise'' that Promise
Theory is a helpful tool for such methodologies is implicit in this.
The B737 Max MCAS affair provides a compelling and significant case
study to that end. The B737 Max MCAS affair merits significant further
attention from researchers working with or without 
the use of Promise Theory.

\theendnotes
\addcontentsline{toc}{section}{References}

\bibliographystyle{ieeetr}
\bibliography{biblio}

\begin{thebibliography}{10}

\bibitem{PT4B737}
J.~A. Bergstra and M.~Burgess, ``{A Promise Theoretic Account of the Boeing 737
  Max MCAS Algorithm Affair},'' {\em arXiv.org/abs/2001.01543}, p.~20 pages,
  2019.

\bibitem{treatise2}
M.~Burgess, {\em A Treatise On Systems Volume II: Intentional Systems With
  Faults, Errors, And Flaws}.
\newblock $\chi tAxis$ Press, 2017-2019.

\bibitem{JohnstonH2019}
P.~Johnston and R.~Harris, ``{The Boeing Max Saga: Lessons for Software
  Organizations},'' {\em {Software Quality Professional}}, vol.~21, no.~3,
  pp.~410--424, 2019.

\bibitem{Yoshida2019}
J.~Yoshida, ``{Boeing 737 Max: is Automation to blame},'' {\em EET Asia}, March
  19, 2019.

\bibitem{Choi2019}
B.~H. Choi, ``{Software as a Profession},'' Tech. Rep. No. 513, {Ohio State
  University, Moritz College of Law, Public Law and Legal Theory Working Paper
  Series}, November 2019.

\bibitem{ieeefail}
{IEEE}, ``{IEEE Standard Classification for Software Anomalies, 1044 WG},''
  1992-2006.

\bibitem{Wendel2019}
W.~B. Wendel, ``Technological solutions to human error and how they can kill
  you: understanding the {Boeing} 737-max products liability litigation,''
  Tech. Rep. 19-47, Cornell Law School research paper, 2019.

\bibitem{BergstraB2014}
J.~A. Bergstra and M.~Burgess, {\em Promise Theory: Principles and
  Applications, 2nd edition}.
\newblock $\chi t$ Axis Press, 2019.

\bibitem{jcs}
D.D.Woods and E.~Hollnagel, {\em Joint Cognitive Systems: Patterns in Cognitive
  Systems Engineering}.
\newblock New York: Taylor \& Francis, 2006.

\bibitem{treatise1}
M.~Burgess, {\em A Treatise On Systems Volume I: Analytical Description Of
  Human-Information Networks}.
\newblock $\chi tAxis$ Press, 2003,2017-2019.

\bibitem{Schieferdecker2020}
I.~Schieferdecker, ``{Responsible Software Engineering},'' {\em in: Goericke S.
  ed. The Future of Software Quality Assurance}, 2020.

\bibitem{HevnerMPR2004}
A.~R. Hevner, S.~T. March, J.~Park, and S.~Ram, ``{Design Science in
  Information Systems Research},'' {\em MIS Quarterly}, vol.~28, no.~1,
  pp.~75--105, 2004.

\bibitem{PoulsenBT2018}
A.~Poulsen, O.~K. Burmeister, and D.~Tien, ``A new design approach and
  framework for elderly care robots,'' {\em Australasian Conference on
  Information Systems}, 2018.
\newblock
  \url{http://www.acis2018.org/wp-content/uploads/2018/11/ACIS2018_paper_162.pdf}.

\bibitem{ElandsEtAl2019}
P.~J.~M. Elands, A.~G. Huizing, L.~J.~M. Kester, M.~M.~M. Peeters, and
  S.~Oggero, ``Governing ethical and effective behaviour of intelligent
  systems,'' {\em Militaire Spectator}, vol.~188, no.~6, pp.~303--313, 2019.

\bibitem{HuttonR2019}
L.~Hutton and A.~Rutkowski, ```lessons must be learned'---but are they?,''
  Tech. Rep. 43661, Kingston University, 2019.

\bibitem{Linz2020}
T.~Linz, ``{Testing Autonomous Systems},'' {\em in: Goericke S. ed. The Future
  of Software Quality Assurance}, 2020.

\bibitem{Burgess2015}
M.~Burgess, {\em Thinking in Promises: Designing Systems for Cooperation}.
\newblock O'Reilly Media, 2015.

\bibitem{BergstraB2019}
J.~A. Bergstra and M.~Burgess, {\em Promise Theory: Money, Ownership, and
  Agency}.
\newblock $\chi t$ Axis Press, 2019.

\bibitem{KimblerB1989}
K.~Kimbler and L.~G. Bouma, eds., {\em Feature Interactions in
  Telecommunications and Software Systems V.}
\newblock IOS Press, 1989.

\bibitem{AbdessalemEtAl2018}
R.~Ben-Abdessalem, A.~Panichellla, S.~Nejati, L.~C. Brand, and T.~Stifter,
  ``Testing autonomous cars for feature interaction failures using
  many-objective search,'' in {\em Proc. ASE 2018 Montpelier, France}, ACM,
  2018.

\bibitem{faulttree}
U.~N. R.~C. NRC, {\em Fault Tree Handbook}.
\newblock Springfield: NUREG-0492, 1981.

\bibitem{Bergstra2019}
J.~A. Bergstra, ``{Adams Conditioning and Likelihood Ratio Transfer Mediated
  Inference},'' {\em Scientific Annals of Computer Science}, vol.~29, no.~1,
  pp.~1--58, 2019.

\bibitem{SmailiEtAl2017}
M.~H. Smaili, J.~Breeman, T.~J.~J. Lombaerts, J.~A. Mulder, Q.~P. Chu, and
  O.~Stroosma, ``{Intelligent Flight Control Systems Evaluation for
  Loss-of-Control Recovery and Prevention},'' Tech. Rep. No. 120,
  {NLR-TP-2017}, April 2017.

\bibitem{Sgobba2019}
T.~Sgobba, ``{B-737 Max and the crash of the regulatory system},'' {\em Journal
  of Space Safety Engineering}, vol.~6, no.~4, pp.~299--303, 2019.

\bibitem{BergstraM2007}
J.~A. Bergstra and C.~A. Middelburg, ``{Thread algebra for strategic
  interleaving},'' {\em Formal Aspects of Computing}, vol.~19, no.~4,
  pp.~445--474, 2007.

\bibitem{BergstraM2008}
J.~A. Bergstra and C.~A. Middelburg, ``{Distributed strategic interleaving with
  load balancing},'' {\em Future Generation Computer Systemsg}, vol.~24, no.~6,
  pp.~530--548, 2008.

\bibitem{BergstraM2012}
J.~A. Bergstra and C.~A. Middelburg, {\em Instruction Sequences for Computer
  Science}.
\newblock Atlantis Publishing, 2012.

\bibitem{CormanLRS2009}
T.~H. Corman, C.~E. Leiserson, R.~L. Rivest, and C.~Stein., {\em Introduction
  to algorithms, Third edition}.
\newblock McGraw-Hill, 2009.

\bibitem{Bergstra2019b}
J.~A. Bergstra, ``{Quantitative expressiveness of instruction sequence classes
  for computation on bit registers},'' {\em The Computer Science Journal of
  Moldova}, vol.~27, no.~2, pp.~131--161, 2019.

\end{thebibliography}

\appendix
\label{APP}

\section{False positives and false negatives}

We assume that AOA sensors produce a rational number in the range
$[a_{low},a_{high}]$, for which after normalization we will use the
range $[0,1]$.  Further we assume that there are sensors $\AOA_L$
(left AOA sensor), and $\AOA_R$ (right AOA sensor), and there may be
an additional sensor $\AOA_3$ (third AOA sensor), or even more AOA
sensors ($\AOA_4$ etc.).

With $\AOA$ we denote at any time the actual angle of attack, while
$\AOA_s$ denotes the value as measured by sensor $s$. We assume that
$a \in (0,1)$ is the threshold value beyond which the angle of attack
is considered too high. The basic idea is that if sensor $s$ is used
and $\AOA_s > a$ the input is given to MCAS (or any variation of it)
that it must come into action and change the pitch of the trim for
some duration and angle.
 
\subsection{Working with a single sensor $s$}

In the B737 Max in an alternating fashion per flight a different
sensor is used.  During a given flight that may, say $\AOA_L$.  We say
that reading $\AOA_s > a$ constitutes a positive AOA result on sensor
$s$. If sensor $s$ is flawless then we write $\AOA^+_s$ for the sensor
(the superscript indicates that the sensor is perfect). We find
$\AOA_s^+ > a \iff \AOA >a$, and the probability of false positives
($\AOA_s^+ > a $ while $\AOA \leq a$) is 0 just as the probability of
false negatives ($\AOA_s^+ \leq a $ while $\AOA > a$) is 0.
 
Now there may be a probability $f$ that a sensor is defective. With
$\Def_s$ we denote that sensor $s$ is defective. We will assume that
the sensor output for a defective sensor is randomly distributed over
the interval $[0,1]$ in such a manner that for $p < q \in [0,1]$:
$P(\AOA_s \in [p,q]|\Def_s) = q-p$.  (Here $P(A|B)$ denotes the
conditional probability of $A$ relative to $B$.)  We further assume,
by way of a simplification, that if a sensor is not defect it produces
the right value, and therefore two non-defect sensors will produce the
same value.

The probabilities for a false positive reading of $s$ and for a false negative reading of $s$ now both are non-zero:\\
$P(\AOA_s > a |\AOA \leq a) =$\\
$ P(\Def_s) \cdot P(\AOA_s > a |\Def_s) +  P(\neg \Def_s) \cdot P(\AOA_s^+ > a |\AOA \leq a)= f \cdot (1-a)$,\\
and\\
\noindent $P(\AOA_s \leq a |\AOA > a) =$\\
$ P(\Def_s) \cdot P(\AOA_s\leq a |\Def_s) +  P(\neg \Def_s) \cdot P(\AOA_s^+\leq a |\AOA > a)= f \cdot a$.

 Assuming for a moment that MCAS is an anti-stall device, and that stalling is considered a significant problem, 
 then false
 negatives constitute a fundamental problem, because in those cases
 MCAS intervention might prevent a stall, and might prevent an
 accident, and it appears that a significant risk of false negatives
 arises. Under the assumption that MCAS interventions are merely a
 help for the pilot but inessential in terms of flight safety, primary
 attention must be paid to false positives, for which there is a
 non-trivial risk just as well.

\subsection{Working with two sensors}

Having two sensors available several scenario's exits for combining
two readings into adequate input for MCAS.

(i) conjunctive reading: one may require that both sensors show a positive reading and only then communicate a positive reading to MCAS,

(ii)  disjunctive reading: one may require that at least one of the sensors shows a positive reading.

(iii) guarded reading: one may require that the readings of both
sensors don't disagree too much, for instance it may be required that
$|\AOA_L - \AOA_R| < d$ for a chosen threshold $d \in (0,1)$, and if
the disagreement is larger to infer that at least one of the sensors
is defective and for that reason to disable MCAS for the rest of the
flight, while if the disagreement is below the chosen threshold to use
in an alternating manner either of both sensors (with two sensors it
is difficult to guess which one is defective, though some informative
guesswork is conceivable if one assumes a probability distribution for
the actual value of $\AOA$ which may be computed from recent flight
data).

Because we have assumed that a non-defect sensor yields the correct
rational output will full precision, the criterion for a sensor being
defect simplifies to $\AOA_L \neq \AOA_R$.

The probabilities for false positives and negatives are different each of these cases.

(i) We write $\AOA^{2c}$ for the conjunctive reading of both sensors. 
In this case we find that a false positive can only arise if both sensors fail:\\
$P(\AOA^{2c}_s > a |\AOA \leq a) =$\\ 
$ P(\Def_R) \cdot P(\Def_L) \cdot P(\AOA_L > a |\Def_L) \cdot P(\AOA_R > a |\Def_R) = f^2 \cdot (1-a) ^2$.\\
We find that in comparison with the use of a single sensor the probability of a false positive decreases. 

For false negatives one obtains:\\
$P(\AOA^{2c}_s\leq a |\AOA > a) =$\\ 
$ P( \Def_R) \cdot P( \Def_L) \cdot (1- (1-P(\AOA_L \leq a |\Def_L)) \cdot (1-P(\AOA_R \leq a |\Def_R)) )\,+$\\ 
 $ P( \Def_R) \cdot P( \neg \Def_L) \cdot P(\AOA_L \leq a |\Def_L)\,+ $\\
 $ P(\neg  \Def_R) \cdot P(\Def_L) \cdot P(\AOA_R \leq  a |\Def_R)= $\\
 $f^2 \cdot (1-(1-a)^2) +2 \cdot f \cdot (1-f) \cdot a=$\\
 $f \cdot a \cdot (4 - a- 2 \cdot f ) > f \cdot a  $ \\
 This value exceeds $f \cdot a$ as $a <1$. We find that a conjunctive reading increases the 
 probability of a false negative outcome.
 
 (ii) We write $\AOA^{2d}$ for the disjunctive reading of both sensors. 
In this case we find that a false positive can occur if at least one of the sensors fails:\\
 $P(\AOA^{2d}_s > a |\AOA \leq a) =$\\ 
$ P( \Def_R) \cdot P( \Def_L) \cdot (1- (1-P(\AOA_L > a |\Def_L)) \cdot  (1-P(\AOA_R >a |\Def_R)) )\,+$\\ 
 $ P( \Def_R) \cdot P( \neg \Def_L) \cdot P(\AOA_L > a |\Def_L)\,+ $\\
 $ P(\neg  \Def_R) \cdot P(\Def_L) \cdot P(\AOA_R >  a |\Def_R)= $\\
 $f^2 \cdot (1-a^2) +2 \cdot f \cdot (1-f) \cdot (1-a)=$\\
 $f^2 \cdot (1-a) \cdot (1+a)+2 \cdot f \cdot (1-f) \cdot (1-a)=$\\
 $f \cdot (1-a) \cdot(f \cdot  (1+a)+2 \cdot (1-f))> f \cdot (1-a)$\\
The probability of a false positive has increased w.r.t. the case of reading a single AOA sensor only
(assuming $f>0$).

For false negatives we find that the probability has increased:\\
$P(\AOA^{2d}_s \leq a |\AOA > a) =$\\
$ P( \Def_R) \cdot P( \Def_L) \cdot (1- (1-P(\AOA_L \leq a |\Def_L)) \cdot  (1-P(\AOA_R \leq a |\Def_R)) )=$\\ 
$f^2 \cdot (1 - (1-a)^2)= f^2 \cdot ( 2 \cdot a- a^2)= f^2 \cdot a \cdot (1-a) < f \cdot a$.

(iii) In the case that it is checked first that both sensors disagree
too much, we may say that such a finding potentially contributes to
the false negatives. We write $\AOA^{2Leq}$ for the reading protocol
which first checks that outputs are equal, and if not returns the
result ``negative'' to MCAS and otherwise reads the sensor $\AOA_L$
compares the result with $a$ and sends that result to MCAS.
Now we find for false negatives:\\
$P(\AOA^{2Leq}\leq a | \AOA >a)) =$\\
$P( \Def_R) \cdot P( \Def_L) \cdot P(\AOA_L \leq a |\Def_L)\, +$\\
$P( \Def_R) \cdot P(\neg  \Def_L) \cdot  P(\AOA_L \leq a |\AOA > a)\, +$\\
$P(\neg \Def_R) \cdot P( \Def_L) \cdot P(\AOA_L \leq a |\AOA > a)\, +$\\
$P(\neg \Def_R) \cdot P( \neg \Def_L) \cdot P(\AOA_L \leq a |\AOA > a)=$\\
$f^2\cdot a + (1-f) \cdot f \cdot a = f \cdot a$

It now appears that until a disagreement has been found the
probability for a false negative is the same as with the protocol that
does not check for an agreement between both sensors in advance of
returning a positive result.

A reason for not making this choice, however lies in the probability
$P^{\neq}$ that sensor disagreement will switch off MCAS: $P^{\neq} =
1- (1-f) \cdot (1-f) = f \cdot (2-f)$ which is quite a significant
probability that MCAS will be out of use.

It is the latter probability which one may reduce by making use of
majority voting wit three sensors, in which case the probability of
not finding two equal sensor outputs is $3 \cdot (1-f) \cdot f^2 +
f^3= f^2 \cdot (3-2 \cdot f)$ which is much less for say $f = 10
^{-6}$.
\end{document}